\documentclass[twocolumn,showpacs,amsmath,amssymb,superscriptaddress]{revtex4}
\usepackage{graphicx}
\usepackage{graphics}
\usepackage{dcolumn}
\usepackage{bm}

\begin{document}
\title{Measuring the condensate fraction of rapidly rotating trapped boson
systems: off-diagonal order from the density profile} 
\author{Jairo Sinova}
\affiliation{Department of Physics, University of Texas at Austin,
Austin TX 78712-1081}
\author{C.B. Hanna}
\affiliation{ Department of Physics, Boise State University,
Boise, Idaho 83725-1570}
\author{A.H. MacDonald}
\affiliation{Department of Physics, University of Texas at Austin,
Austin TX 78712-1081}
\date{\today}

\begin{abstract}

We demonstrate a direct connection between the density profile of a system
of ultra-cold trapped bosonic particles in the rapid-rotation limit and its 
condensate fraction.  This connection can be used to probe the 
crossover from condensed vortex-lattice states to uncondensed quantum fluid
states that occurs in rapidly rotating boson systems as the particle density
decreases or the rotation frequency increases.  We illustrate our proposal
with a series of examples, including ones based on models of realistic finite
trap systems, and comment on its application to freely expanding boson 
density profile measurements.

\end{abstract}

\pacs{PACS numbers: 03.75.Fi, 05.30.Jp, 73.43.Nq}

\maketitle 

When a system of coherent trapped bosons is brought to equilibrium in a
rotating frame of reference, vortices are induced in the
order parameter of its condensate and centrifugal forces lead to 
radial expansion.  The influence of the rotating frame may
be described mathematically by introducing an effective magnetic field
oriented along the axis of rotation and reducing the
radial confinement strength.
Taking this point of view, very rapidly rotating boson systems 
enter the quasi-two-dimensional (2D) strong-field quantum Hall regime 
in which interactions are always important.  Many-fermion systems in 
the quantum Hall regime have been thoroughly studied over the past 
twenty years and have exhibited a wide variety of subtle and novel
behaviors, many of which were completely unanticipated by theory \cite{qhebook}. 
This history provides a powerful motivation for creating
boson quantum-Hall systems.
Most current experiments in rotating Bose-Einstein systems 
create condensates in which the effective magnetic field is
relatively weak and axial-direction quantum confinement effects are 
negligible when compared to interaction strengths.
For such condensates, the nucleation of vortices and the
formation of vortex lattices, together with their decay and evolution,
have been observed experimentally and studied theoretically by several
groups \cite{madison00,aboshaeer01,haljan01,hodby02,aboshaeer02,engles02,
fetter01}. 

The limit of quasi-2D bosons in the rapid-rotation
(or equivalently, quantum Hall) regime has so far
been studied only theoretically
\cite{sinova02,castin99,cooper99,cooper01,peredes01, ho01}.
The physics of trapped bosons in the quantum Hall regime
is enriched by the near-degeneracy of single-particle energy levels,
which enhances the importance of condensate quantum fluctuations
\cite{sinova02}, reduces the condensate fraction, and eventually, at low
boson densities, leads to quantum fluid states \cite{cooper01,peredes01}
that are expected to have unusual properties analogous to those that have been
uncovered in studies of the fermion quantum Hall effect.
One of the major stumbling blocks in contemplating studies of 
bosons in the quantum Hall regime has been the lack of a reliable tool 
to directly measure the degree of Bose-Einstein condensation and 
the accompanying long-range many-boson phase coherence. 
Unlike its familiar zero-rotation cousin, a vortex-lattice condensate
in the quantum Hall regime
does not have a characteristic velocity distribution that allows it to be
identified from the {\em velocity} distribution of the released system.
In this paper, we show that in the quasi-2D rapid-rotation limit there is,
however, a surprisingly simple relationship between diagonal elements of the
many-boson density-matrix and the off-diagonal order responsible for coherence.
The condensate
fraction can be probed simply by measuring the spatial distribution of the
boson {\em density}.

Before explaining the simple but powerful relation that we propose be exploited,
we first discuss its regime of validity.
A system of $N$ interacting bosons of mass $M$
in a cylindrical trap (with radial and axial trap frequencies
$\Omega_r$ and $\Omega_z$) that is rotating with angular velocity
$\Omega {\bf \hat{z}}$ is well-described by the rotating-frame
Hamiltonian \cite{leggett01}
\begin{eqnarray}
\cal{H}&=&\sum_{i=1}^N \left\{ \frac{({\bf p}_i -
M\Omega {\bf \hat{z}}\times {\bf r}_i)^2}{2M}
+\frac{M}{2} \left[(\Omega_r^2-\Omega^2)(x_i^2+y_i^2) \right. \right.
\nonumber\\&& \left. \left.
+ \Omega_z^2z_i^2 \right] \right\} +\sum_{i<j=1}^N V({\bf r}_i-{\bf r}_j) ,
\label{hamiltonian}
\end{eqnarray}
where $V({\bf r})$=$g\delta({\bf r})$ is a hard-core interaction potential
of strength $g = 4 \pi \hbar^2 a_s/M$,
and $a_s$ is the $s$-wave scattering length.
This Hamiltonian is equivalent to that of a system of charge-$Q$ bosons
with weakened radial confinement, under the influence of a
magnetic field ${\bf B}=(2M\Omega/Q){\bf\hat{z}}$, with cyclotron frequency
$\Omega_c=QB/M = 2 \Omega$ and magnetic length
$\ell=\sqrt{\hbar/(2 M\Omega)}$.   
The Hilbert space of this system is spanned by the
axial ($z$-direction) quantum states and the various Landau-level states
produced by the effective magnetic field.
For $\Omega$ close to $\Omega_r$, the effective confinement potential
becomes very weak, which causes the bosons to spread out radially and,
in the limit of interest to us here (see below), allows all bosons to be 
accomodated in the lowest-energy axial quantum state.  

In the quasi-2D limit,
the boson system can be described by a 2D model whose short-range
interactions have strength $g_{2D}= 
\hbar^2 a_s\sqrt{8\pi}/M l_z$, where 
$l_z=\sqrt{\hbar/M\Omega_z}$. 
The areal density of bosons in such a trap can be estimated using a 
2D Thomas-Fermi (TF) approximation in which
$M (\Omega_r^2-\Omega^2)r^2/2 + g_{2D} n_{2D}$ 
is constant inside the system \cite{note1}
Since the bosons can form rather uniform-density clouds (of the type
favored by short-range interactions) entirely within the
LLL, there is little incentive for them to occupy higher
kinetic-energy Landau levels, unless the typical interaction strength
$g_{2D} n_{2D}$ is much larger than the Landau-level splitting 
$2 \hbar \Omega$.  The TF approximation
boson Landau-level filling factor is given by
$ \nu(r)\equiv 2\pi \ell^2 n_{2D}(r) = \nu_0 (1 - r^2/R^2)$, where 
$R^2 = \sqrt{4 g_{2D} N/ \pi M (\Omega_r^2-\Omega^2)}$, and
$\nu_0^2
={\sqrt{2\pi} l_z N(\Omega_r^2/\Omega^2-1)}/{4 a_s}$.
The parameter 
\begin{equation}
\gamma_{LL} \equiv 
\frac{g_{2D} n_{2D}(r=0)}{2 \hbar \Omega}
= \sqrt{\frac{(\Omega_r^2/\Omega^2-1) a_s N}{l_z \sqrt{2 \pi}}}.
\label{eq:figofmerit} 
\end{equation}
characterizes the degree of Landau-level mixing induced by interactions.
Neglecting boson Landau-level mixing, as we do in the rest of the paper,
is a good approximation provided that $\gamma_{LL}$ is not large.
Similarly, neglecting mixing of higher axial energy eigenstates is a good approximation
provided that $\gamma_{z} = 2 \gamma_{LL} \Omega/\Omega_z$ is not too
large.  In mean-field theory, the effect of a finite value of $\gamma_{z}$
is to weaken axial confinement at the center of the trap and therefore to
effectively weaken $g_{2D}$.
The conclusions we reach below would not be altered by a
position-dependent effective interaction strength.
We therefore expect our results to apply for values of $\gamma_{LL}$ 
and $\gamma_{z}$ that are substantially larger than one,
although more detailed calculations will be required to establish the
range of the rapid-rotation limit more quantitatively.
Values of $\gamma_{LL}$ and
$\gamma_{z}$ that are not much larger than unity are already within reach
of current experimental designs\cite{cornell02}.

Condensation of a system of trapped bosons (we consider here only the
zero-temperature case) is discussed most generally in terms  
of the spectrum of its one-particle density matrix (ODM),
whose position representation
is related to the many-boson wave function by 
\begin{eqnarray}
\rho_1({\bf r},{\bf r}')\equiv N \int d{\bf r}_2\cdots d{\bf r}_N
\Psi_N^*({\bf r},{\bf r}_2,\cdots,{\bf r}_N)\nonumber\\\times
\Psi_N({\bf r}',{\bf r}_2,\cdots,{\bf r}_N)=
\langle\hat{\psi}^\dagger({\bf r})
\hat{\psi}({\bf r}')\rangle ,
\label{dm}
\end{eqnarray}
where $\rho_1({\bf r},{\bf r}) = n({\bf r})$ is the boson density,
and the trace of the ODM is the total particle number, $N$.
If one or more of the eigenvalues, $\lambda_i$, of the ODM is extensive
(i.e., $\lambda_i\propto N$), then the system is
Bose-condensed \cite{leggett01}.
In a particle-conserving system, the ratio
\begin{equation}
\beta=\frac{{\rm Tr} (\rho_1^2)}
           {({\rm Tr} \rho_1)^2}
\end{equation}
will be of order unity \textit{only} if the system is Bose-condensed
and the condensate fraction will be $\sqrt{\beta}$,
up to a correction of order $1/N$.

The many-body physics of quasi-2D boson systems in a magnetic field
simplifies in the rapid-rotation limit because the many-boson Hilbert
space can be projected to the lowest Landau level (LLL).
These simplifications are often most conveniently
captured in the symmetric gauge, where they imply that many-particle wave
functions must have the form
$\Psi_N({\bf r}_1,{\bf r}_2,\cdots,{\bf r}_N) = f[z]
\exp\left(-\sum_k |z_k|^2/4\ell^2\right)$ ,
where $f[z]$ is analytic in each of the $N$ complex bosonic coordinates 
$z_k=x_k+iy_k$.  The analyticity property has played a key role in achieving
an understanding of many-fermion physics in the quantum Hall regime,
for example in Laughlin's recognition \cite{rbl:1983} that incompressible
states would occur at certain Landau-level filling factors.  
Because of this property, the ODM is 
completely specified by the functional form of its diagonal matrix element.
For fermions, this property leads \cite{macdonald88} to Hartree-Fock exchange 
energies that depend only on the particle density; for bosons, it has more
fundamental implications, as we now discuss.

It follows from of $\Psi_N({\bf r}_1,{\bf r}_2,\cdots,{\bf r}_N)$ that 
\begin{equation}
\rho_1({\bf r}',{\bf r})= n[z^*,z']
\exp\left(-\frac{|z|^2}{4 \ell^2}
-\frac{|z'|^2}{4 \ell^2}+\frac{z {z'}^*}{2\ell^2}\right) ,
\label{eq:odm}
\end{equation}
where $n[z^*,z]$ is the boson density expressed as a function of
$z=x+iy$ and $z^*=x-iy$.
Equation~(\ref{eq:odm}) can be used to express $\beta$ in terms of the
2D Fourier transform of the boson density:
$n[{\bf r}]=n[z^*,z]
=\int {d \bf q} n({\bf q})\exp[i(q^*z+q z^*)/2]/(4 \pi^2)$ ,
where  and $q=q_x+iq_y$.
We find that 
\begin{equation}
\beta=\frac{\ell^2}{2\pi N^2}\int d{\bf q}
 |n({\bf q})|^2 e^{q^2\ell^2/2}.
\label{eq:beta} 
\end{equation}
We propose that Eq.~(\ref{eq:beta}) be used to measure the condensate
fraction of rapidly rotating quasi-2D Bose-Einstein condensates. 

\begin{figure}
\hskip -0.18 in
\includegraphics[width=2.8in,height=2.8in]{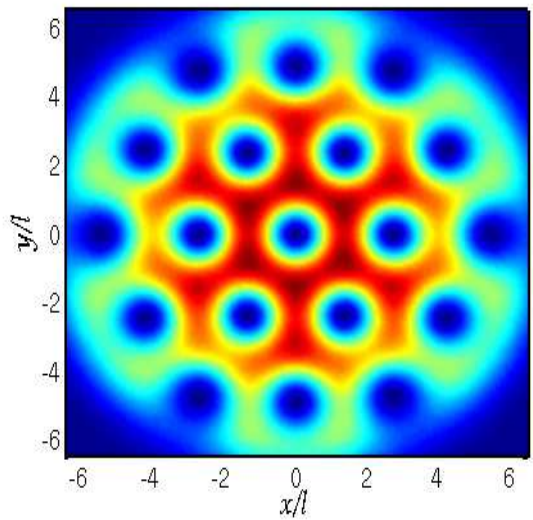}
\includegraphics[width=3.2in]{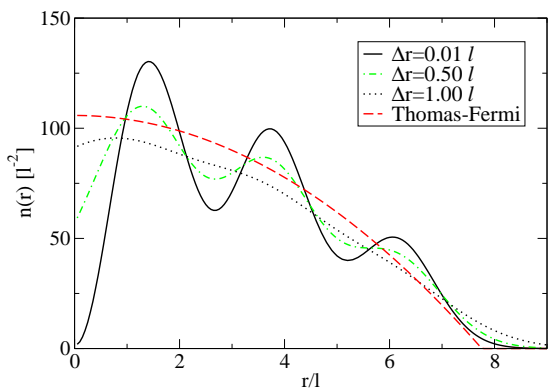}
\caption{\label{figure1} Upper figure:
Density profile for the ground state of a rapidly rotating boson system
with $\Omega_r/\Omega=1.04$, $g_{2D} M /\hbar^2 =0.0058$, and $N=10,000$.
These parameters are similar to ones that can be achieved by current
experimental systems.  For these parameters,
the optimal number of vortices is 19.
Lower figure:
Angle-averaged boson filling-factor profile with radial broadening $\Delta r$. 
The long-dashed line in this figure is the TF approximation boson
filling-factor profile for the same parameters.
}
\end{figure}
\begin{figure}
\hskip -0.2 in
\includegraphics[width=2.8in,height=2.8in]{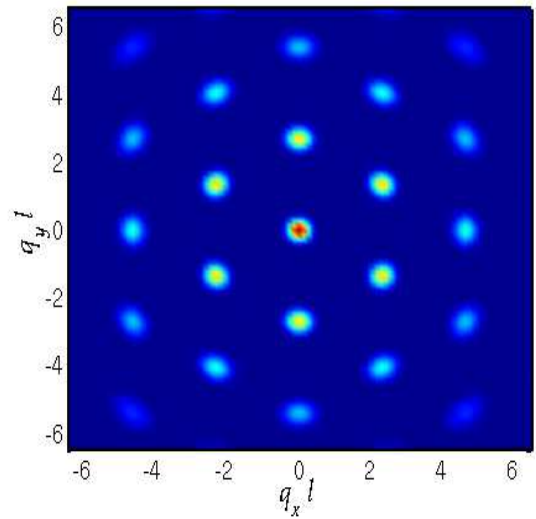}
\caption{\label{figure2} $|n(q)|^2 \exp(q^2\ell^2/2)$ for the density
profile shown in Fig.~\ref{figure1}.  The axes are those of the 2D
wave vector, multiplied by the effective magnetic length, $\ell$.
}
\end{figure}

To understand the strengths and weaknesses of this condensate measure
as an experimental probe, it is helpful to discuss several different examples.  
We first consider the case of a rapidly rotating fully condensed BEC that
contains a single vortex; for typical traps and interaction strengths, such
a state would occur only for boson particle numbers that are too small
to study at the present time.  In this case, the Gross-Pitaevskii (GP)
order parameter \cite{leggett01} is
$\phi(z) = \sqrt{{N}/({4\pi\ell^4})} z \exp(-|z|^2/4\ell^2)$
which leads to $n(q) = N (1-q^2\ell^2/2) \exp(-q^2\ell^2/2)$,
and yields $\beta=1$, as expected.

A qualitatively similar but more realistic case is illustrated in
Fig.~\ref{figure1}. 
The upper panel of this figure shows the mean-field-theory 2D boson
density profile, obtained by solving the GP equations
following Refs. \onlinecite{castin99} and \onlinecite{feder01},
while allowing the trap's rotational symmetry to be broken
and optimizing with respect to the number of vortices, for the parameters
$\Omega_r/\Omega=1.04$, $g_{2D}/(\hbar^2 M)=0.0058$, and $N=10,000$.
These parameters were chosen to closely match those of
current experimental systems \cite{note2}.
The lower graph in Fig.~\ref{figure1} shows the profile of the angle-averaged
boson filling factor.  Under these circumstances,
we expect \cite{sinova02} that the condensate
fraction will be large, at least in the limit of zero temperature. 
When smoothed radially over a length $ \sim \ell$, the boson density profile
agrees reasonably well with that predicted by TF theory,
illustrating the dominant role of interactions in determining the
course-grained density profile, even in the rapid-rotation limit.
In Fig.~\ref{figure2}, we show the integrand of Eq.~(\ref{eq:beta})
that is involved in calculating $\beta$, for the configuration
shown in Fig.~\ref{figure1}.  
In practice however, only the first few rings of Fig. ~\ref{figure2} can be
resolved in current experiments and a finite wave-vector cutoff 
must be introduced \cite{cornell02}.  In this case,
two different estimates of $\beta$ might prove useful. 
A lower bound $\beta_-$ is obviously obtained by calculating $\beta$ 
using an experimentally constrained momentum cutoff $q_{\rm max}$.
A separate estimate of $\beta$, which is more likely to be useful in practice
for current experimental systems, is motivated by observing that 
in the case of a quasi-2D (untrapped) condensate each Brillouin-zone in momentum space contributes
equally to $\int d^2 {\bf q} |n({\bf q})|^2 \exp (|q^2|\ell^2/2)$.
It follows that the ratio of diffraction peak strengths 
integrated over Brillouin-zones centered on ${\bf q} \ne 0$ (condensate contributions only) and 
${\bf q} = 0$ (condensate plus normal fluid contributions)  
can be used to estimate $\beta$.  The efficacy of this approach could
be checked by comparing estimates from different shells \cite{note3}.  
For the example of Fig.~\ref{figure2}, we obtain
$\beta_- \approx 0.27, 0.57, 0.83$ for cutoffs of
$q_{\rm max}\ell = 4, 6, 9$, and, using the first shell for the second estimate 
we obtain $\beta \approx 0.93$; 
i.e., all show clear evidence of Bose-Einstein condensation, 
albeit with an inaccurate value of the condensate fraction. 
(The condensate fraction is, of course, always one for GP approximation density profiles.) 
As this calculation illustrates, we 
do not anticipate that our method will be especially useful for
measuring small deviations of the condensate fraction from unity in
strongly condensed systems.
We do, however, expect that meaningful measurements of $\beta$
will be possible when the condensate fraction is strongly reduced due to either
thermal or quantum fluctuations, and anticipate that strong reductions will
occur at available temperatures even at very high values of $\nu_0$.

We next examine the quantum fluid regime in which rotational symmetry
is not broken and the 2D density profile should be given approximately
by the TF approximation expression, 
$n(r)=\nu_0(1-r^2/R^2)/(2\pi\ell^2)$.
For $R/\ell \gg 1$,
we find that $\beta \approx (2/3)\nu_0/N$,
after introducing an ultraviolet cutoff to remove
an artificial cusp in the TF density.
This result illustrates the ability of our condensate measure
to capture the property that, in the rapid-rotation limit,
BEC can occur only when accompanied by broken rotational symmetry.
Incompressible boson states \cite{cooper01,peredes01,ho01}
at very small filling factors are expected to produce more uniform
density profiles, but they also have $\beta\sim\nu/N$ and are thus
uncondensed.

As a final example, we consider a state with two independent condensates,
A and B, with a density profile
$n({\bf r})=\lambda_A n_A({\bf r})+\lambda_B n_B({\bf r})$,
where $\lambda_A+\lambda_B=1$.
For the case in which species A is in an uncondensed constant-density state
and species B is a large vortex-lattice condensed state,
we obtain $\beta=\lambda_B^2+(2\lambda_A-\lambda_A^2)\nu/N$.
Therefore,
$\beta$ will be of order unity
even if only one of the species is condensed.

Another experimental issue that might sometimes arise in the measurement of $\beta$
is the determination of the boson density profile from  
free-expansion images.
As shown in Refs. \onlinecite{castin99} and \onlinecite{kagan96},
the evolution of the order parameter under 2D free expansion
(i.e., when the confinement in the $z$-direction is maintained)
is given by
$\phi_{\rm lab}({\bf r},t) = 
\exp[-i(\Omega L_z-\mu)\tau(t)/\hbar]
\phi_{\rm rot}({\bf r}/\lambda(t))/\lambda(t)$ ,
for a system obeying the time-dependent GP equation.
Here $\phi_{\rm lab}$ is the order parameter viewed in the laboratory frame,
and $\phi_{\rm rot}({\bf r})$ is the order parameter before
the trap has been turned
off (obtained from the time-independent GP equation in the initial
rotating frame), $\lambda(t)=\sqrt{1+\Omega_r^2 t^2}$, and
$\Omega_r\tau(t)=\arctan(\Omega_r t)$.
The only assumption used to obtain this result, aside from
ignoring interactions  with the thermal cloud
(which is justified by the short expansion times involved),
is the validity of the time-dependent GP equation.
In the quantum Hall regime bosons can enter a 
regime in which quantum fluctuations render these assumptions invalid,  
particularly when regions of the 
condensate near the edge that have low local filling fractions become important.
It is possible that such low-density regions
could evolve under free expansion differently than 
the higher filling fraction regions near the center.  
Hence, whenever $\beta$ is no longer of order unity,
the analysis of the boson density profile using free-expansion 
measurements is suspect and one must resort to non-destructive
({\it i.e.} unexpanded) measurements of the density profile. Given the
large radial expansion expected in this regime of interest here, however, 
it may be possible to resolve a sufficiently large number of 
shells of reciprocal space $n(\bf q)$ diffraction peaks to apply the  
method we propose.  

In summary, we propose a new experimental tool to explore the transition
between a vortex-lattice Bose-Einstein condensate and an
uncondensed quantum-fluid state in rapidly rotating boson systems.
This new tool derives from a deep connection between the density profile
of a system of ultra-cold bosonic particles in the rapid-rotation limit
and their degree of condensation.
It applies in the quasi-2D rapid-rotation limit that current experiments
are approaching. 

\begin{acknowledgments}
The authors gratefully acknowledge helpful interactions with E.A. Cornell
P. Engels and K.W. Madison.
This work was supported by the Welch Foundation and by the National Science
Foundation under grants DMR-0115947, DMR-9972332, DMR-0206681, and EPS-0132626.
\end{acknowledgments}


\begin{thebibliography}{23}
\bibitem[{not({\natexlab{a}})}]{note1}
\bibinfo{note}{Because of the proportionality between angular momentum and
  $\langle r^2 \rangle$ in the LLL, the TF approximation corresponds to a flat
  Landau band, with the mean-field single-particle energy of occupied orbitals
  independent of angular momentum. Condensates which mix higher Landau levels
  can have more uniform density and lower interaction energy because they can
  have smaller vortex cores, but only at a substantial cost in kinetic energy.}
\bibitem[{not({\natexlab{c}})}]{note3}
\bibinfo{note}{Quantum fluctuations will presumably have successively larger
  effect on more remote diffraction peak shells.}
\end{thebibliography}

\end{document}